\documentclass[times]{aa501}
\usepackage{psfig}
\usepackage{natbib}
\bibpunct{(}{)}{;}{a}{}{,}

\begin{document}

\def\eps{\varepsilon}
\def\aap{A\&A}
\def\apj{ApJ}
\def\apjl{ApJL}
\def\mnras{MNRAS}
\def\aj{AJ}
\def\nat{Nature}
\def\aaps{A\&A Supp.}
\def\bub{{\rm b}}
\def\cl{{\rm cl}}
\def\rp{{\rm rp}}
\def\st{\sigma_{\rm T}}
\def\e{{\rm e}}
\def\c{{\rm c}}
\def\p{{\rm p}}
\def\me{m_\e}
\def\lesssim{\mathrel{\hbox{\rlap{\hbox{\lower4pt\hbox{$\sim$}}}\hbox{$<$}}}}
\def\gtrsim{\mathrel{\hbox{\rlap{\hbox{\lower4pt\hbox{$\sim$}}}\hbox{$>$}}}}

\title{Radio and X-Ray Detectability of Buoyant Radio Plasma Bubbles in
Clusters of Galaxies}
\titlerunning{Detectability of Buoyant Radio Bubbles in
Clusters of Galaxies}
\author{Torsten A. En{\ss}lin \and Sebastian Heinz}
\authorrunning{T. A. En{\ss}lin \and S. Heinz}
\institute{Max-Planck-Institut f\"{u}r
Astrophysik, Karl-Schwarzschild-Str.1, Postfach 1317, 85741 Garching,
Germany} 
\date{today}

\abstract{The {\em Chandra} X-ray Observatory is finding a surprisingly
large number of cavities in the X-ray emitting intracluster medium (ICM),
produced by the release of radio plasma from active galactic nuclei.  In
this Letter, we present simple analytic formulae for the evolution of the
X-ray deficit and for the radio spectrum of a buoyantly rising bubble. The
aim of this work is to provide a theoretical framework for the planning and
the analysis of X-ray and radio observations of galaxy clusters.  We show
that the cluster volume tested for the presence of cavities by X-ray
observations is a strongly rising function of the sensitivity.
 } \maketitle

\section{Introduction\label{sec:intro}}
An early key result from the {\em Chandra} X-ray observatory was the
discovery of numerous X-ray cavities in clusters of galaxies, preceded
by pioneering detections by the {\it ROSAT} satellite.  For example,
cavities were found in the Perseus cluster
\citep{1993MNRAS.264L..25B}, the Cygnus-A cluster
\citep{1994MNRAS.270..173C,2000MNRAS.318L..65F}, the Hydra-A cluster
\citep{2000ApJ...534L.135M}, Abell 2597 \citep{2001ApJ...562L.149M},
Abell 4059 \citep{1998ApJ...496..728H,heinz2002}, Abell 2199
\citep{fabian2001moriond}, Abell 2052 \citep{2001ApJ...558L..15B},
close to M84 in the Virgo Cluster \citep{2001ApJ...547L.107F}, in the
RBS797 cluster \citep{2001A&A...376L..27S}, and in the MKW3s cluster
\citep{astro-ph0107557}.  In most of these cases, the cavities are
clearly coincident with the lobes of a radio galaxy at the cluster
center.  However, some clusters exhibit also cavities without
detectable radio emission, namely in Perseus, Abell 2597, and Abell
4059. The latter class of cavities are also believed to be filled with
radio plasma, but during the buoyant rise of the very light radio
plasma in the cluster atmosphere
\citep{gull1973,2000A&A...356..788C,2001ApJ...554..261C,2001MNRAS.325..676B,astro-ph0201125}
adiabatic expansion and synchrotron/inverse Compton radiation losses
should have diminished the observable radio emitting electron
population, leaving behind a so called {\it ghost cavity} or {\it
radio ghost} \citep{1999dtrp.conf..275E}. The detectability of an
X-ray cavity should also decrease with increasing distance from the
cluster center due to the decreasing ratio of the missing X-ray
emission from the volume occupied by the bubble to the fore- and
background X-ray emission.

\section{Rising buoyant bubbles}

While the early phase of radio galaxy evolution is characterized by
supersonic expansion into the surrounding medium, the radio lobes
quickly settle into pressure equilibrium with the ICM after the AGN
has shut off.  Our description sets in at this moment $t_1$, where the
bubble is located at a cluster radius $r_1$ with volume $V_{\bub,1}$
and pressure $P_1 = P_{\rm ICM}(r_{1})$.  The bubble will then quickly
approach a terminal velocity $v_{\rm b}(r)$, governed by the balance
of buoyancy and drag forces.  During its rise, the bubble volume
changes according to the adiabatic law $V_\bub(r) = V_{\bub,1}\,
(P(r)/P_1)^{-1/\gamma_\rp}$, where the adiabatic index $\gamma_\rp$ is
close to $4/3$, which we will take as our fiducial value for numerical
examples.  The magnetic field strength should evolve according to
$B(t) = B_1\,(V_\bub(t)/V_{\bub,1})^{-2/3} =
B_1\,(P(t)/P_1)^{2/(3\,\gamma_\rp)}$, if the expansion of the bubble
is isotropic.  For simplicity, we only consider spherical bubbles with
radius $r_\bub$, which gives sufficiently accurate estimates for most
applications. If the bubble becomes highly deformed or even
disintegrates, more sophisticated models than ours will have to be
used. We further assume, that entrainment of environmental gas into
the bubble is dynamically insignificant on the considered time-scales,
implying that the bubble is X-ray dark. Numerical simulations
\citep[e.g.,][]{2001ApJ...549L.179R} support the latter.

It is often convenient to express physical quantities like the bubble
volume and the magnetic field strength in terms of the values they
would have if the bubble were adiabatically moved to the cluster
center. We denote these by the subscript\ $0$ (e.g., $V_{\bub,0}=
V_\bub(r=0) = V_{\bub,1}\, (P_0/P_1)^{-1/\gamma_\rp}$). As our working
example, we will investigate a cluster described by an isothermal
$\beta$-profile with a density profile of $\rho(r) =
\rho_0\,(1+(r/r_\c)^2)^{-3\,\beta/2}$, pressure $P(r) =
P_0\,(1+(r/r_\c)^2)^{-3\,\beta/2}$, and constant sound speed $c_{\rm
s}\delta$.

We define the origin of the cluster coordinate system at the cluster
center, with the $x$ and $y$ axis defining the image plane and the
$z$-axis the line of sight to the observer, and the coordinate system
oriented so that the bubble center is located in the $x$-$z$ plane at
$\vec{r} = (r\,\cos\theta,0,r\,\sin\theta)$. Its projected distance
from the cluster center is $R = \mu\,r$, where $\mu = \cos\theta$. The
angle $\theta$ should be roughly conserved along the bubble's
trajectory in a spherical cluster atmosphere.

The buoyancy speed of the bubble can be estimated from the balance of
buoyancy and drag forces. The buoyancy force
\begin{equation}
F_{\rm buoyancy} = \frac{4}{3} \,\pi\, r_\bub^3\,g\,\rho = \frac{4}{3}
\,\pi\, r_\bub^3\,\frac{dP}{dr} 
\end{equation}
depends on the gravitational acceleration $g$, which can be expressed
via the pressure gradient in a cluster with hydrostatical
equilibrium. The hydrodynamical drag of subsonic motion is well
approximated by
\begin{equation}
\label{eq:fdrag}
F_{\rm drag} = C_{\rm d}\,\pi\,r_\bub^2 \,\rho \,v^2,
\end{equation}
with $C_{\rm d} \approx 0.5$. Using $r_\bub = r_{\bub,0}
(P/P_0)^{-1/(3\,\gamma_\rp)}$, the bubble velocity is given by
\begin{equation}
\label{eq:vbub}
v_\bub(r) = \left(\frac{4\,r_\bub}{3\,C_{\rm
d}\,\rho}\,\frac{dP}{dr}\right)^{\frac{1}{2}}\! = \frac{r_\c}{\tau}
\left(\frac{r}{r_\c}\right)^{\frac{1}{2}} \!
\left(1+\frac{r^2}{r_\c^2}\right)^{-\frac{1}{2} +
\frac{\beta}{4\,\gamma_\rp}}
\!\!\!\!\!\!\!\!\!\!\!\!\!\!\!\!\!\!\!\!\!\!,
\end{equation}
where
\begin{equation}
\label{eq:tau}
\tau = \sqrt{\frac{C_{\rm d}\,\gamma_{\rm ICM}\, r_\c}{\beta\,r_{\bub,0}}}
\,\frac{r_\c}{2\,c_{\rm s}} \approx \frac{1}{2}\,\sqrt{\frac{r_\c}{r_{\bub,0}}}
\frac{r_\c}{c_{\rm s}}
\end{equation}
is the characteristic buoyancy timescale across one core radius. The
typical rise velocity is a fraction ($\propto \sqrt{r_{\bub}/r_\c}$)
of the sound speed $c_{\rm s}$. The time the bubble needs to travel
from $r_1$ to $r_2$ is then
\begin{equation}
t(r_1,r_2) = \frac{\tau}{2}\, \left[{\rm B}_{\frac{x^2}{1+x^2}}
\left(\frac{1}{4}, \frac{\beta}{4\,\gamma_\rp}-\frac{3}{4} \right)
\right]_{r_1/r_\c}^{r_2/r_\c},
\end{equation}
where ${\rm B}_x(a,b) \equiv \int_{0}^{x}dy\, y^a\,(1 - y)^b$ denotes the
incomplete beta function, and $[f(x)]_{x_1}^{x_2} \equiv f(x_2) - f(x_1)$.

If the initial bubble is large compared to the cluster core
Eq. \ref{eq:vbub} can give supersonic rise velocities. In such a case
Eq. \ref{eq:fdrag} is no longer valid. Instead, strongly increased
dissipation will limit the rise velocity to the subsonic regime. In
such a case one will adopt $v_\bub = \alpha_\bub\,c_{\rm s}$, with
$\alpha_\bub \approx 0.5$.

\section{X-ray deficit evolution}

The X-ray emissivity at cluster radius $r$ and for a given density
and pressure profile is $\epsilon_{\nu} = \Lambda_\nu(T(r))\,
{n(r)}^2$, where $n$ is the electron number density and $\Lambda_{\nu}
(T)$ is the plasma cooling function.  

In the following we will assume that the gas surrounding the bubble has
essentially settled back into hydrostatic equilibrium in the dark matter
potential, which is unperturbed by the radio galaxy activity.

In general, the surface brightness integral must be solved
numerically, but in the case of a $\beta$-model atmosphere, it can be
expressed in closed form.  For an undisturbed line of sight (i.e., not
intersecting a bubble) the observed surface brightness (neglecting
cosmological corrections) is given by
\begin{equation}
\label{eq:unpertprof}
	I_{\rm X,\cl}(x,y) = {I_0}\, \left( 1+ \frac{x^2 +
	y^2}{r_\c^2} \right)^{-3\beta + \frac{1}{2}} + I_{\rm bg},
\end{equation}
where $I_0 = \Lambda_{\rm T}\, n_0^2 \, r_\c\, (4\pi)^{-1}\,{\rm B}(3\beta
-\frac{1}{2},\frac{1}{2})$ is the central, and $I_{\rm bg}$ the background
photon flux. ${\rm B}(a,b)$ is the beta function $B_{x=1}(a,b)$.

For a line of sight intersecting the surface of a bubble, the intersection
points are located at $(x,y,z_{\pm})$, with
\begin{equation}
	z_{\pm} = r\,\sqrt{1-\mu^2} \pm \sqrt{r_\bub^2 - y^2 - (x - r
	\,\mu)^2}.
\end{equation}
The surface brightness in the region occupied by the bubble is given
by
\begin{eqnarray}
	I_{{\rm X,\bub}}(x,y) &=& I_{\rm X,\cl}(x,y) -{I_0}\, \left( 1+ \frac{x^2 +
	y^2}{r_\c^2}\right)^{-3\beta + \frac{1}{2}} \\
	&&\times \left[\frac{{\rm sgn}(z)}{2}\,\frac{{\rm
	B}_{\varpi(z)}(\frac{1}{2}, 3\beta - \frac{1}{2})}{{\rm
	B}(3\beta - \frac{1}{2}, \frac{1}{2})}\right]_{z_{-}}^{z_{+}}
	 \nonumber
\end{eqnarray}
where we defined $\varpi(z) \equiv z^2/(r_\c^2+x^2+y^2+z^2)$.

Another important diagnostic is the number of missing counts $\Delta
N_{\rm X}$ from the cavities, i.e., the photons emitted from a
spherical region of radius $r_{\rm b}$ at distance $r$ from the center
in an unperturbed cluster atmosphere during the exposure interval
$t_{\rm obs}$.  For a detector with effective area $A_\nu$ and a
source distance of $D$ we define
\begin{equation}
\lambda_{\rm T} = \int \!\! d\nu\, \frac{A_\nu\,\Lambda_\nu (T)\,t_{\rm
obs}}{4\pi D^2 h\,\nu}.
\end{equation}
Then,
\begin{equation}
	\Delta N_{\rm X} = \frac{\pi \lambda_{\rm T} n_0^2 r_{\rm
	c}^3}{6\beta-2}\left[ {\rm B}_{\frac{1}{v}}(3\beta
	-\frac{3}{2},\frac{1}{2}) +
	\frac{v^{2-3\beta}\,r_\c}{(3\beta-2)\,r} \right]_{v_{-}}^{v_{+}}
\end{equation}
where $v_{\pm} = 1 + [(r_{\rm b} \pm r)/r_{\rm c}]^2$.  

\begin{figure}[t]
\begin{center}
\psfig{figure=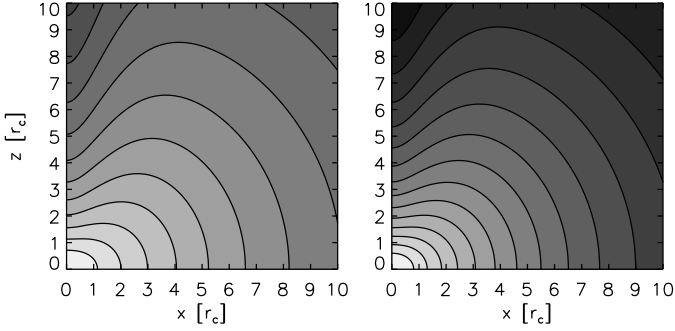,width=0.5 \textwidth,angle=0}
\end{center}
\vspace{-0.8cm}
\caption[]{\label{fig:sigcont}Contours of the X-ray deficit
significance $({\mathcal S}/{\mathcal N})$ of a bubble in a galaxy
cluster. The contours mark locations at which a bubble has a
significance which is lower by powers of 2 than its significance if
located at the cluster center (in the lower left corner of each
figure, vertical axis is parallel to the line of sight). Left: the
bubble volume expands adiabatically (with $\gamma_\rp=4/3$) with the
pressure of the isothermal cluster ($\beta= 3/4$). Right: the bubble
volume is assumed to be independent of location
($\gamma_\rp\rightarrow \infty$). In both figures, an X-ray background
with 1/100 of the central cluster surface brightness is assumed
($\delta = 0.01$).}
\end{figure}

In order to design searches for cavities in promising cluster
candidates, it is useful to estimate the expected signal to noise
ratio (${\mathcal S}/{\mathcal N}$) of a potential observation.  The
signal is the expected number of missing photons from the volume
occupied by the bubble, ${\mathcal S} ={\Delta N}_{\rm X}$. The
background, on top of which the missing photons have to be detected,
is the expected number of photons ${N}_{\rm exp}$ from the area on the
sky subtended by the cavity, assuming there were no cavity. This
assumes that the instrument point spread function is not more extended
than the cavity.  The noise is then given by the fluctuation in this
photon number, which is ${\mathcal N} \approx \sqrt{{N}_{\rm exp}}$.

\begin{figure}[t]
\begin{center}
\psfig{figure=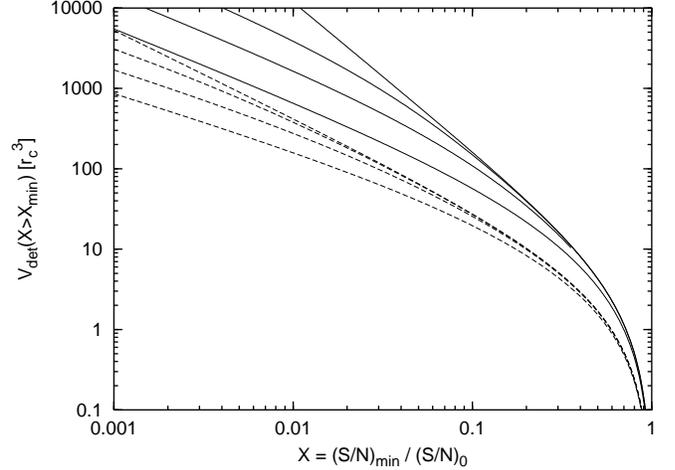,width=0.5\textwidth,angle=0}
\end{center}
\vspace{-0.8cm}
\caption[]{\label{fig:volX}Volume of the cluster [in units of $r_{\rm
c}^3$] within which the signal to noise ratio $({\mathcal S}/{\mathcal
N})$ of the X-ray deficiency of a radio bubble exceeds a minimum, for
the detection required value $({\mathcal S}/{\mathcal N})_{\rm
min}$. The latter is given in units of $({\mathcal S}/{\mathcal
N})_0$, the signal to noise of a comparable bubble moved to the
cluster center.  The solid lines are for an adiabatically expanding
bubble ($\gamma_\rp = 4/3$), and the dashed lines for an
incompressible bubble ($\gamma_\rp \rightarrow \infty$). The
background to central cluster X-ray surface brightness ratio takes the
values $\delta =$ 0, 0.001, 0.01, and 0.1 from top to bottom. The
(top) curves with $\delta = 0$ are given by Eq. \ref{eq:Vdetana}. We
adopted $\beta= 3/4$.}
\end{figure}

The signal-to-noise ratio of a small bubble (so that the cluster density
does not vary significantly across its diameter, $r_{\rm b} \ll {\rm
Max}[r_{\rm c}, r]$), in an isothermal $\beta$-model is well
approximated by
\begin{equation}
\label{eq:SN1}
\frac{\mathcal S}{\mathcal N} \!=\!
\frac{4 r_{\bub,0}^2\,n_0\,(1\!+\!  r^2/r_\c^2)^{-3\beta
+\frac{\beta}{\gamma_\rp}}}{3[(1\!+\!\mu^2\,r^2/r_\c^2)^{-3 \beta +
\frac{1}{2}} \!+\! \delta]^\frac{1}{2}}
\!\!\left[\frac{ \pi\, \lambda_{\rm T}}{r_\c\, {\rm
B}(3\beta\!-\!\frac{1}{2},\frac{1}{2})}\right]^\frac{1}{2}\!\!\!\!,
\end{equation}
where $\delta$ is the ratio of the uniform background to central
cluster photon flux. In most applications $\delta \ll 1$.  We write
\begin{equation}
\left(\frac{\mathcal S}{\mathcal N}\right)_0 = \frac{\mathcal S}{\mathcal
N} (r=0) = \frac{4\, n_0 \,r_{\bub,0}^2}{3\,\sqrt{1+\delta}}
\left(\frac{\pi\,\lambda_T}{r_\c\, {\rm
B}(3\beta-\frac{1}{2},\frac{1}{2})} \right)^\frac{1}{2} 
\end{equation}
for the signal to noise ratio of a cavity moved to the cluster
center. We write $({\mathcal S}/{\mathcal N})_{\rm min}$ for the
minimum signal to noise required for a firm bubble detection. We
define $X = ({\mathcal S}/{\mathcal N})_{\rm min}/({\mathcal
S}/{\mathcal N})_0$, which is $<1$ for all detectable bubbles. The
subvolume of the cluster within which a bubble of the chosen size can
be detected is shown in Fig. \ref{fig:sigcont}. The dependence of the
total integrated cluster volume $V_{\rm det}$, at which such a bubble
can be detected, is displayed in Fig. \ref{fig:volX} as a function of
$X$ and $\delta$. For $\delta = 0$ the volume integral can be
estimated analytically, and yields
\begin{equation}
\label{eq:Vdetana}
V_{\rm det} = \frac{4}{3}\,\pi\, (X^{-\xi}\, Y)^{\frac{3}{2}}\,r_\c^3 +
2\,\pi\, \eta\, X^{-\psi}\,
B_Y(\frac{3}{2},\frac{3}{2} \eta)\,r_\c^3,
\label{eq:signaltonoise}
\end{equation}
where $Y= 1-X^{\xi}$, $\xi = 1/(3\beta-\beta/\gamma_\rp)$, $\psi =
6/(6\beta+1-4\beta/\gamma_\rp)$, and $\eta=
({6\beta-1})/({6\beta+1-4\beta/\gamma_\rp})$.  

If the question of interest is the detectability of cavities of a given
fixed size at all cluster radii one can use the incompressible limit
$\gamma_\rp \rightarrow \infty$ of Eqs. \ref{eq:SN1} and
\ref{eq:signaltonoise}. This case is also displayed in
Figs. \ref{fig:sigcont} and \ref{fig:volX}.

\section{Radio spectrum evolution}

Synchrotron, inverse Compton, and adiabatic losses cool an electron with
initial dimensionless momentum $p_1 = \mbox{momentum}/({\rm m_\e\,c})$ down
to
$p(p_1,t) =  p_1/( p_1\,q(t) + (V_\bub(t)/V_{\bub,1})^{\frac{1}{3}})$,
where
\begin{equation}
\label{eq:q2}
q = \frac{1}{p_{\rm max}} = a_0 \int_{r_1}^{r_2}\!\!\!dr\,\frac{u_B(r) + u_{\rm
C}}{v_\bub(r)}\,\left(\frac{P(r)}{P(r_2)}
\right)^{\frac{1}{3\,\gamma_\rp}}
\end{equation}
is the inverse maximal momentum.  Here, $a_{0} = \frac{4}{3}\,
\sigma_{\rm T}/(m_\e\, c)$. The magnetic and photon (mostly CMB)
energy densities are denoted by $u_B=B^2/(8\,\pi)= u_{B,0}\,
(P/P_0)^{4/(3\,\gamma_\rp)}$ and $u_C$ respectively. We have re-written the time
integral in \cite{2001A&A...366...26E} as an integral over the
trajectory of the bubble from $r_1$ to $r_2$.  For our isothermal
cluster this integral evaluates to
\begin{eqnarray}
\label{eq:qiso}
q &\!=\!& \frac{a_0\,\tau}{2}\!
\left(1+\frac{r_2^2}{r_\c^2}\right)^{\frac{\beta}{2\,\gamma_\rp}}
\!\!\!\!\!\!\!\!\!\!\!\! \nonumber \\ & & \times \left[ u_{B,0}\,{\rm B}_{\frac{x^2}{1+x^2}}
(\kappa,\zeta_B) + u_{\rm C}\, {\rm B}_{\frac{x^2}{1+x^2}} (\kappa,\zeta_{\rm C})
\right]_{r_1/r_\c}^{r_2/r_\c} \!\!\!\!\!\!\!
\end{eqnarray}
with $\zeta_B= (11 \beta-3\gamma_\rp)/(4\gamma_\rp)$, $\zeta_{\rm C} =
(3\beta-3\gamma_\rp)/(4\gamma_\rp)$, $\kappa = 1/4$, and $\tau$ given by
Eq. \ref{eq:tau}.

If a bubble is rising with a constant velocity $v_\bub$
(e.g. because it approaches the sound velocity) the following parameters
give the correct $q$ value: $\zeta_B= (5\beta-\gamma_\rp)/(2\gamma_\rp)$,
$\zeta_{\rm C} = (\beta-\gamma_\rp)/(2\gamma_\rp)$, $\kappa = 1/2$, and
$\tau= r_\c/v_\bub$.

An initially relativistic power-law electron spectrum $f(p,t_1)\,dp =
f_{0} \, p^{-s}\,dp$ for $p_{\rm min\,1} < p < p_{\rm max\, 1}$
becomes
\begin{equation}
\label{eq:spec2}
f(p,t) = {f}_{0} \, (P(t)/P_1)^{\frac{s -1}{3\gamma_{\rm rp}}}\,
p^{-s}\, \left( 1 - p\,q(t) \right)^{s -2}
\end{equation}
for $p_{\rm min}(t)= p(p_{\rm min\,1},t) < p < p_{\rm max}(t) =
p(p_{\rm max\, 1},t)$.  The synchrotron luminosity of these electrons
is
\begin{equation}
\label{eq:lnu1}
L_\nu(t) = \int_0^\infty \!\!\!\!dp\,P_\nu(p,B)\,f(p,t)\,,
\end{equation}
where $P_\nu(p,B)$ is the synchrotron kernel.  In the following we use
the monochromatic approximation, which means that the total
synchrotron radiation losses of an electron ($\dot{E}_\e(p) =
-a_0\,u_B\,p^2$) are emitted at a single characteristic frequency
$\nu(p,B) = \Lambda_{\rm s}\, B\, p^2$ with $\Lambda_{\rm s} = 3
\,e/(2\,\pi\,m_\e\,c)$. In this approximation, the synchrotron
spectrum extends up to $\nu_{\rm max} = \Lambda_{\rm s}\,B_2\,p^2_{\rm
max}(t) \leq \Lambda_{\rm s}\,B_2\,q^{-2}(t)$ and can be written
analytically as
\begin{equation}
\label{eq:lnu2}
L_\nu\! =\! L_{\nu,1}
\left(\frac{P_2}{P_1}\right)^{\!\frac{2s}{3\gamma_\rp}}\!\!
\left(1- \sqrt{\frac{\nu}{\Lambda_{\rm s}\,B_1}}\,
\left(\frac{P_1}{P_2}\right)^\frac{1}{3\gamma_\rp}\, q
\right)^{\!s-2} 
\!\!\!\!\!\!\!\!\!,
\end{equation}
where 
\begin{equation}
L_{\nu,1} = \frac{\st\,c\,B\,f_0}{12\,\pi\,\Lambda_{\rm s}}
\left(\frac{\nu}{\Lambda_{\rm s}\,B_1}\right)^{\!-\frac{s-1}{2}}
\end{equation}
was the original luminosity at $r_1$.  Eq. \ref{eq:lnu2} (or the more
accurate Eq. \ref{eq:lnu1}), and Eq. \ref{eq:qiso} (or
Eq. \ref{eq:q2}) can be used to calculate the optical thin part of the
radio spectrum of the source.

\section{Conclusion}

\begin{figure}[t]
\begin{center}
\psfig{figure=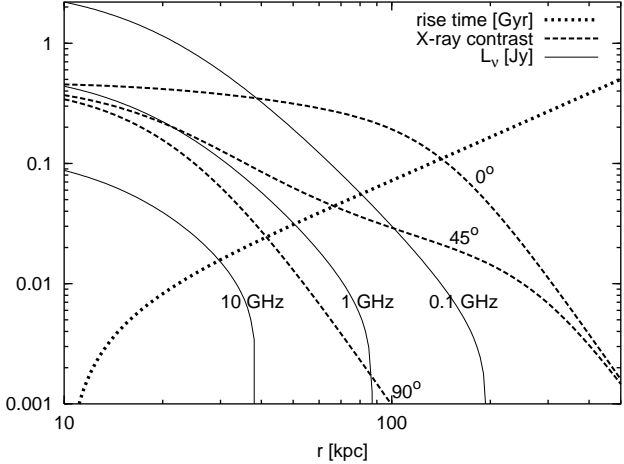,width=0.5\textwidth,angle=0}
\end{center}
\vspace{-0.8cm}
\caption[]{\label{fig:example}Bubble's central X-ray contrast
(compared to the undisturbed cluster) for various angles between plane of
sky and Bubble's trajectory, its radio flux, and its rising time
as a function of the (unprojected) radial position. Adopted
parameters: $r_\c = 20$ kpc, $c_{\rm s} = 1400$ km/s, $\beta = 3/4$,
$r_1 = 10$ kpc, $r_{\bub,1} = 8$ kpc, $B_0 = 10 \mu$G, $\delta =
10^{-3}$, $s = 2.4$, $F_{\rm 1.4 GHz}(r_1) = 0.35$ Jy.}
\end{figure}

As an aid and stimulus for future observations we provided simple
analytic formulae of X-ray and radio properties of rising buoyant
bubbles of radio plasma in galaxy clusters.  The detectability of a
bubble decreases with its cluster radius. The X-ray contrast of a
bubble moving in the plane of the sky decreases slowly until the X-ray
background dominates, then it drops rapidly. The contrast of a bubble
moving along the line of sight declines quickly outside the cluster
core. Similarly, the radio luminosity of the bubble at a given
frequency declines rapidly with increasing cluster radius and suddenly
vanishes when cooling has removed the emitting electrons. After that
point, only a weak flux of synchrotron-self Comptonized emission
remains \citep{EnsslinSunyaevI01}.  Fig. \ref{fig:example} illustrates
these dependencies for an example with parameters similar to Perseus
A.

\begin{acknowledgements}
We acknowledge useful comments by Eugene~Churazov, Federica~Govoni,
Daniel E. Harris, Gopal-Krishna, Brian R. McNamara, and the referee
Paul E.~J. Nulsen.
\end{acknowledgements}



\end{document}